\documentclass[prl,amssymb,amsmath,twocolumn,twoside,showpacs,
preprintnumbers,lengthcheck]{revtex4}

\usepackage{epsfig}
\usepackage{mathptm}

\begin{document}

\preprint{\underline{Phys. Rev. Lett. 89, 268702 (2002); cond-mat/0207409}}

\title{Extreme Self-Organization in Networks Constructed from Gene Expression
Data}

\author{Himanshu \surname{Agrawal}}
\email{feagrawa@wicc.weizmann.ac.il}
\affiliation{Department of Physics of Complex Systems,
	     Weizmann Institute of Science, Rehovot 76100, Israel}

\date{Received 31 July 2002; published 12 December 2002}

\begin{abstract}

We study networks constructed from gene expression data obtained from many types
of cancers. The networks are constructed by connecting vertices that belong to
each others' list of $K$-nearest-neighbors, with $K$ being an \textit{a priori}
selected non-negative integer. We introduce an order parameter for
characterizing the homogeneity of the networks. On minimizing the order
parameter with respect to $K$, degree distribution of the networks shows
power-law behavior in the tails with an exponent of unity. Analysis of the
eigenvalue spectrum of the networks confirms the presence of the power-law and
small-world behavior. We discuss the significance of these findings in the
context of evolutionary biological processes.

\end{abstract}


\maketitle

Recent technical advancements have led to widespread use of gene chips for
quantizing and monitoring the expression level of thousands of genes in parallel
\cite{Land99Gerh99}. Presently, gene expression profiling has become an
important tool for diagnosis and classification of diseases. It is being used
extensively for identifying genes responsible for specific conditions, e.g.,
various cancers \cite{Alon99,Nott01,Golu99,Pero99,Pero00}. This is done using
specialized clustering techniques developed in recent years
\cite{Eis98,Blat96,Braz00}. Gene expression data can also be used for
constructing networks of coexpressed and coregulated genes. Since proteins are
the end product of gene expression, various types of protein networks
\cite{Jeo01Mas02} and gene networks are directly related. Consequently, networks
of coregulated genes are expected to play key role in biological processes. In
this letter we outline results that show the relevance of these networks in
evolutionary biological processes.

The volume of gene expression data obtained from typical experiments is enormous
and contains information on expression of all the genes (presently almost
10\,000 or more) marked on the chip. In any given condition most of the genes
are not important and do not express. As a result, before the expression data
can be used for constructing networks, it requires extensive processing and
filtering to eliminate uninformative genes. We skip these details here as they
can be found with the source of the data
\cite{Alon99,Nott01,Golu99,Pero99,Pero00} and elsewhere \cite{Eis98,Braz00}.
Henceforth, we assume that expression data for the selected set of informative
genes is available in the form of a matrix having $N$ rows and $D$ columns. The
rows represent the genes and the columns represent the samples/tissues.
Furthermore, the expression values of each of the genes in this matrix are
normalized to have a mean of zero and variance of unity across the samples.

The normalized expression levels are treated as coordinates of $N$ genes, that
form the vertices of the networks, in $D$ dimensional space of samples. The
network construction algorithm requires specification of the maximum number of
neighbors $K$, $0 \le K < N$, that a vertex can have. For a given $K$, gene
network is constructed using the following two step procedure. (i) For each
vertex $i$, $i = 1$, \ldots, $N$, make a list $L_{i}$ of its first $K$ nearest
neighbors. (ii) Connect all vertices $i$ and $j$ through an edge if $i \in
L_{j}$ and $j \in L_{i}$, otherwise the vertices are not connected. This
algorithm is derived from the $K$-nearest-neighbor parameter estimation method
\cite{Fuk90}. With some heuristic modifications it has been used earlier in
clustering analysis of various types of data \cite{Blat96}. We used Euclidean
norm as the distance measure for making the list of $K$ nearest neighbors. Other
distance measures can also be used. The results presented herein remain
unaltered as long as the distance measure preserves the ordering of points
obtained from the Euclidean measure.

For a given data set, the topological structure of networks generated by this
algorithm depends strongly on the parameter $K$. For $K = 0$, the network
consists of $N$ isolated vertices, and for $K = N-1$ each vertex is connected to
all the other vertices. For most of the values of $K$, these networks have more
than one connected component. As $K$ increases, the connectivity of each vertex
grows depending on its local environment. Vertices that lie close to each other
tend to get mutually connected in preference to those lying farther away. Thus,
all connected components in these networks have a small-world structure.
Furthermore, for $K \gtrsim 3$ the networks have a giant connected component.

We analyzed networks constructed using several published gene expression data
sets. To ensure that our results are not affected by possible bias of technology
used for manufacturing the gene chips, we used expression data sets obtained
using both oligonucleotide arrays \cite{Alon99,Nott01,Golu99} as well as cDNA
microarrays \cite{Pero99,Pero00}. For each network we calculated $P(z)$ the
probability of finding a vertex of degree $z$. $P(z)$ is normalized by $N$ so
that $\sum_{z}P(z) = 1$. Since these networks are small, $P(z)$ is very noisy.
As a result, we calculated the cumulative probability distribution
\begin{equation}\label{E:FzPz}
  F(z) = \sum_{i = z}^{z_{\text{max}}} P(i)
\end{equation}
where $z_{\text{max}}$ is the maximum degree in the network.

We also define a quantity $c = z+1$. This quantity gives the size of the
smallest cluster around a vertex with connectivity $z$ that includes the vertex
and its neighbors. It can also be considered as the size of a ``droplet'' that
is formed by the vertex and its neighbors. Since, the smallest and the largest
values of $z$ are 0 and $z_{\text{max}}$, the corresponding values for $c$
become 1 and $c_{\text{max}} = 1 + z_{\text{max}}$, where $c_{\text{max}}$ is
the largest droplet size. The probability density $\tilde{P}(c)$ and cumulative
distribution $\tilde{F}(c)$ corresponding to $c$ are defined similarly to those
for $z$. The linear relationship of $c$ and $z$ implies that $\tilde{P}(c) =
P(z)$ and $\tilde{F}(c) = F(z)$. Outside the range of $z$ and $c$, corresponding
probability density functions are defined to be zero. Thus, $\tilde{F}(c)|_{c
\le 1} = F(z)|_{z \le 0} = 1$ and $\tilde{F}(c)|_{c>c_{\text{max}}} =
F(z)|_{z>z_{\text{max}}} = 0$.

The homogeneity of the networks can be characterized by a single order
parameter. A suitable candidate for this is the area $\Lambda$ enclosed by the
$\tilde{F}(c)$ versus $c^{\star} = c/c_{\text{max}}$ curve between $c^{\star} =
0$ and $1$. It takes values in the range $0$ to $1$ depending on the homogeneity
of the network. Since $c_{\text{max}}$ is finite and the values of $c$ are
evenly spaced, $\Lambda(K)$ is easily calculated using Trapezoidal rule and
equals
\begin{equation}\label{E:LP}
  c_{\text{max}} \Lambda(K)
    = \frac{1}{2} \left[ 1 - P(z_{\text{max}}) \right]
      + \sum_{i=0}^{z_{\text{max}}} (i+1) P(i).
\end{equation}
The terms with factor of $1/2$ represent the area of strips at the boundary.
Their contribution vanishes as the number of strips increases. It is zero if
$P(z)$ is a delta function.

From Eq.~(\ref{E:LP}) the value of $c_{\text{max}} \Lambda(K)$ is easily
identified as the mean size $\bar{c} = 1+\bar{z}$ of the droplets, where
$\bar{z}$ is the average degree of vertices in the network. Thus, $\Lambda(K)$
is the average droplet size normalized with the size of the largest droplet in
the network. It measures separation between the mean and maximum droplet sizes
in the network and functions as an indicator of the the overall behavior of
$F(z)$ and $P(z)$. Very small values of $\Lambda(K)$ imply that $F(z)$ descends
sharply from unity to almost zero at a small $z \ll z_{\text{max}}$ and becomes
nearly flat with plateau stretching until $z_{\text{max}}$. The corresponding
$P(z)$ has long tails with weight centered at small $z$. $\Lambda(K)$ close to
unity implies that $F(z)$ stays nearly flat at unity for most of $z \le
z_{\text{max}}$ and descends sharply to zero at some $z \approx z_{\text{max}}$.
In this case $P(z)$ is sharply peaked similar to a delta function near
$z_{\text{max}}$. Intermediate values of $\Lambda(K)$ imply a decaying $F(z)$
corresponding to various forms of $P(z)$ including those that rise slowly to a
peak and then decay slowly via sharply truncated power-law tails.

\begin{figure}
\centerline{\epsfxsize50mm\epsfbox{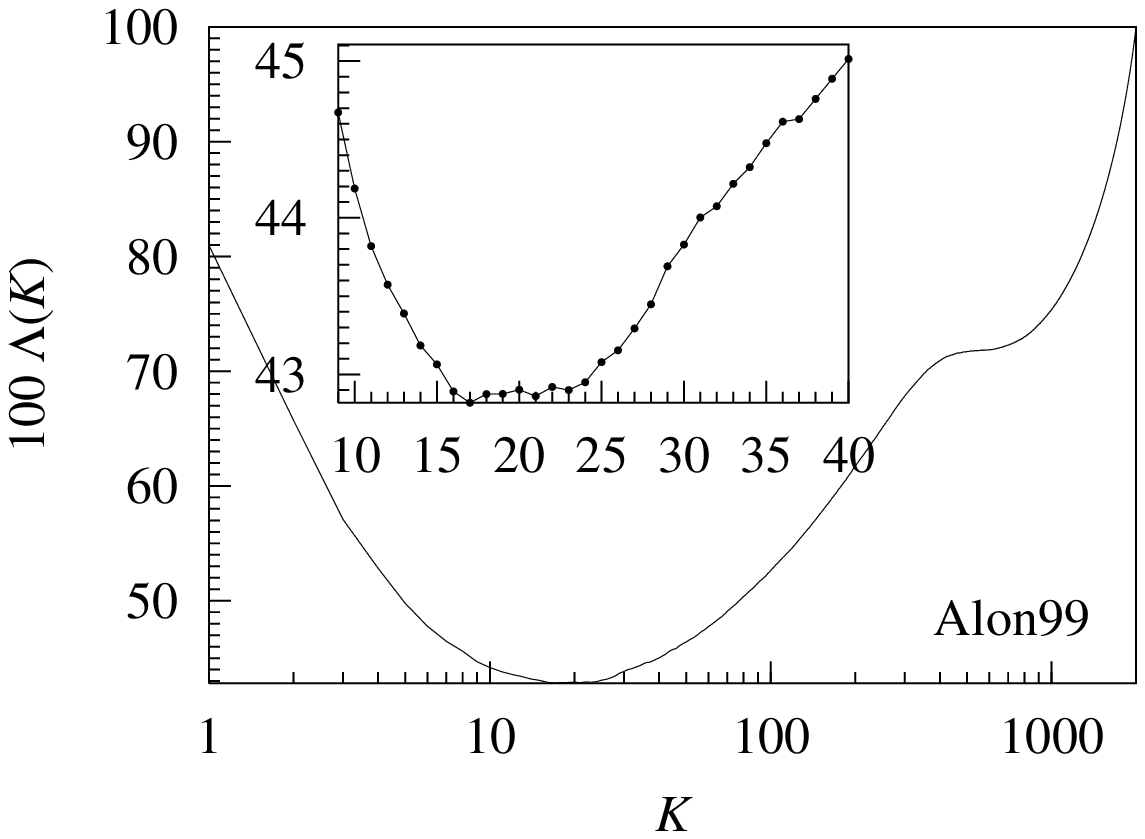}}
\centerline{\epsfxsize50mm\epsfbox{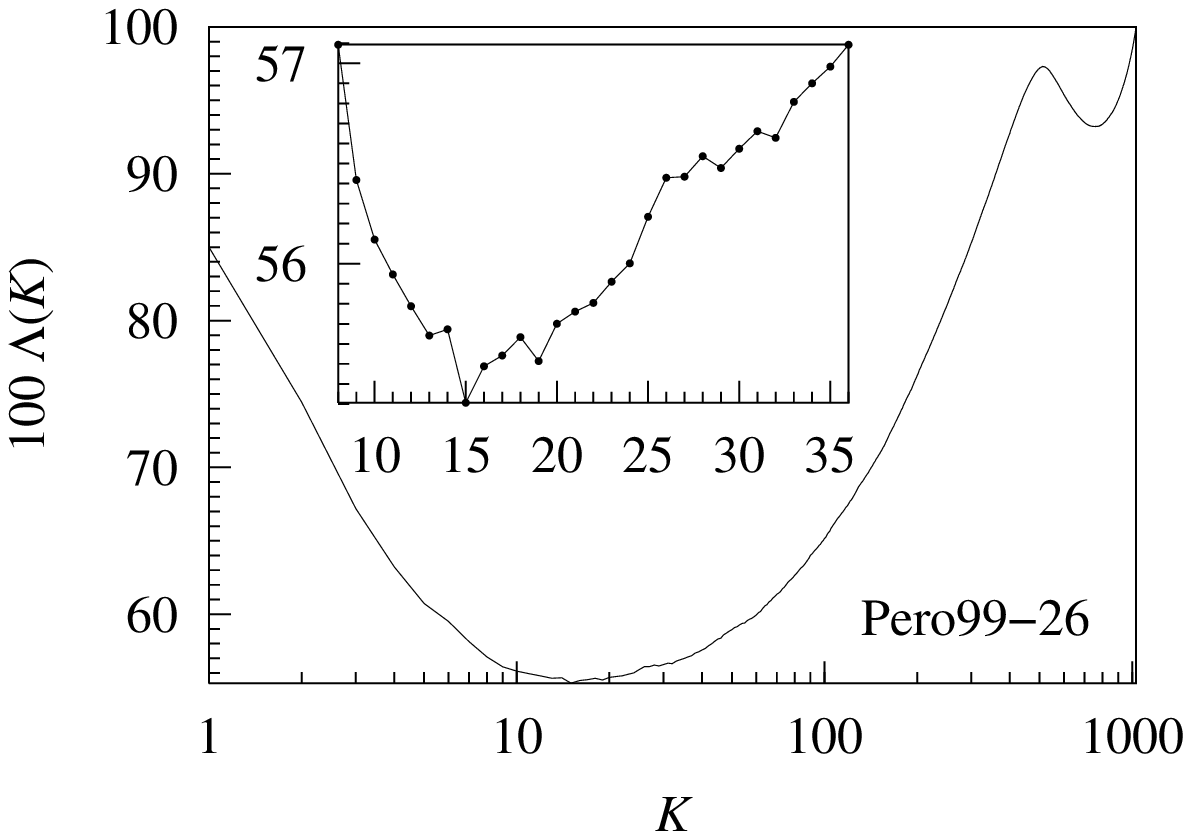}}

\caption{\label{F:LK} Variation of $\Lambda(K)$ with $K$ for networks
constructed from colon cancer data of Alon \textit{et al} \cite{Alon99} and
breast cancer data of Perou \textit{et al} \cite{Pero99}. The insets show blow
up of a small zone around the minima. The keys are as in \cite{DKey}.}

\end{figure}

Figure~\ref{F:LK} (Alon99) shows the behavior of $\Lambda(K)$ for networks
constructed using colon cancer data of Alon \textit{et al.} \cite{Alon99}. The
figure shows, that as $K$ is increased from 1 to $N-1$, $\Lambda(K)$ initially
decreases and attains a minimum at some value of $K = K_{1}$ (here $K_{1}
\approx 16$). The minimum is nearly flat and persists until $K = K_{2}$ (here
$K_{2} \approx 24$). As $K$ is increased beyond $K_{2}$, $\Lambda(K)$ continues
to increase, reaching its maximum value of 1 at $K = N-1$. Fig.~\ref{F:LK}
(Pero99-26) shows similar behavior of $\Lambda(K)$ in networks constructed using
breast cancer data of Perou \textit{et al.} \cite{Pero99} with a second minimum
at $K \approx 3N/4$. The second minimum is usually a finite size effect. It can
also occur if there is high inhomogeneity on length scales large compared to the
nearest neighbors scale. It is very shallow compared to that between $K_{1}$ and
$K_{2}$ because the probability density function corresponding to the size of
``droplets'' on larger length scales are sharper compared to $P(z)$ and also
have smaller tails. The higher order minima, however, are not relevant. Behavior
similar to that seen in Fig.~\ref{F:LK} was observed in networks from several
other gene expression data sets also \cite{Nott01,Golu99,Pero00}.

The behavior of $\Lambda(K)$ divides the networks in three classes. (i) The
networks corresponding to $1 \le K < K_{1}$ have few connections between
vertices but have high homogeneity. (ii) The networks for $K_{1} \le K \le
K_{2}$ are somewhat better connected and highly inhomogeneous. (iii) In the
networks for $K_{2} < K \le N-1$, the vertices have many connections and high
homogeneity that increases with $K$.

\begin{figure*}
\centerline{\epsfxsize55mm\epsfbox{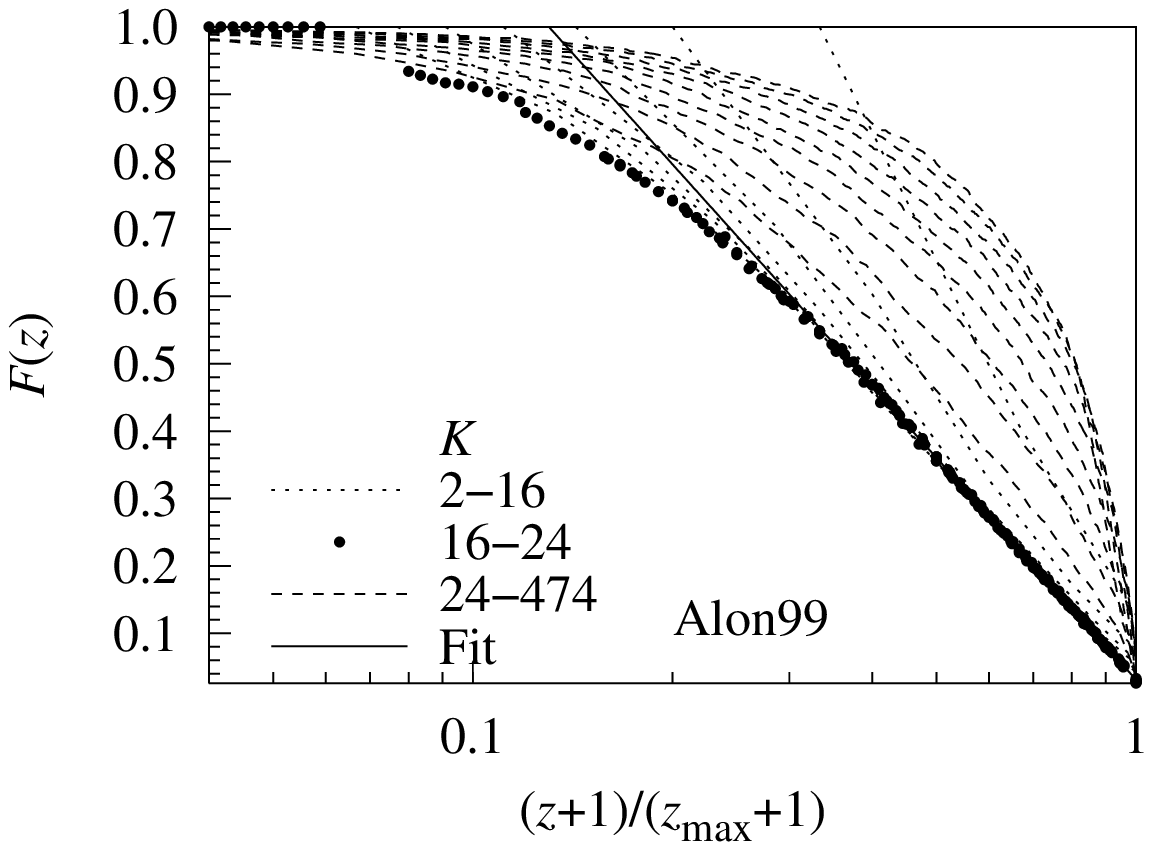}%
	    \epsfxsize55mm\epsfbox{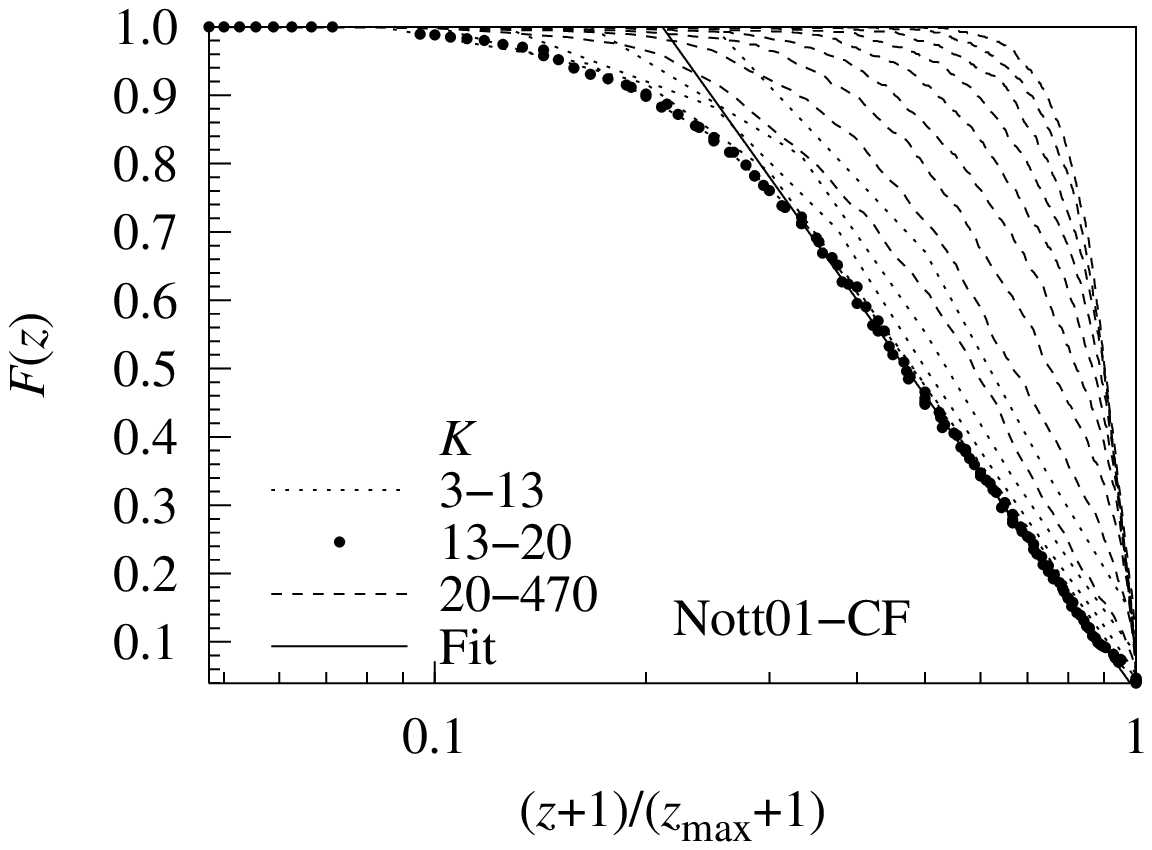}%
	    \epsfxsize55mm\epsfbox{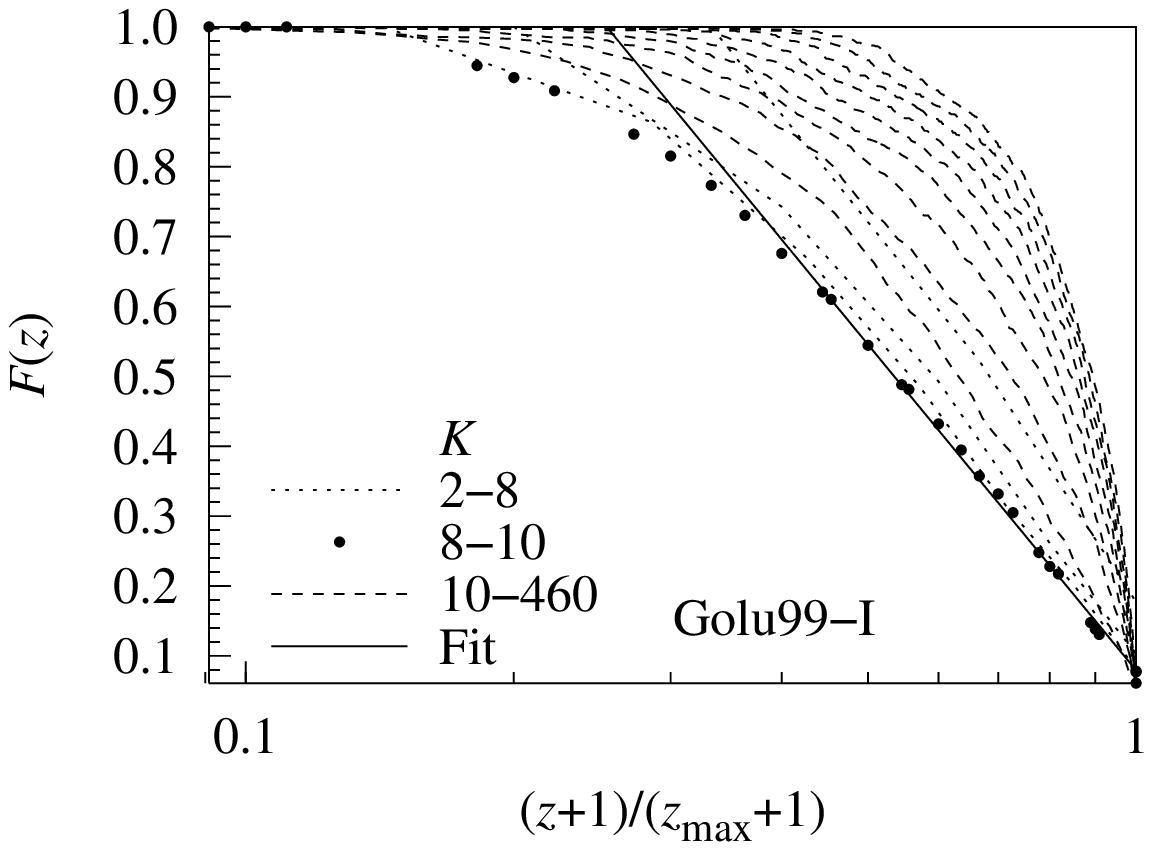}}
\centerline{\epsfxsize55mm\epsfbox{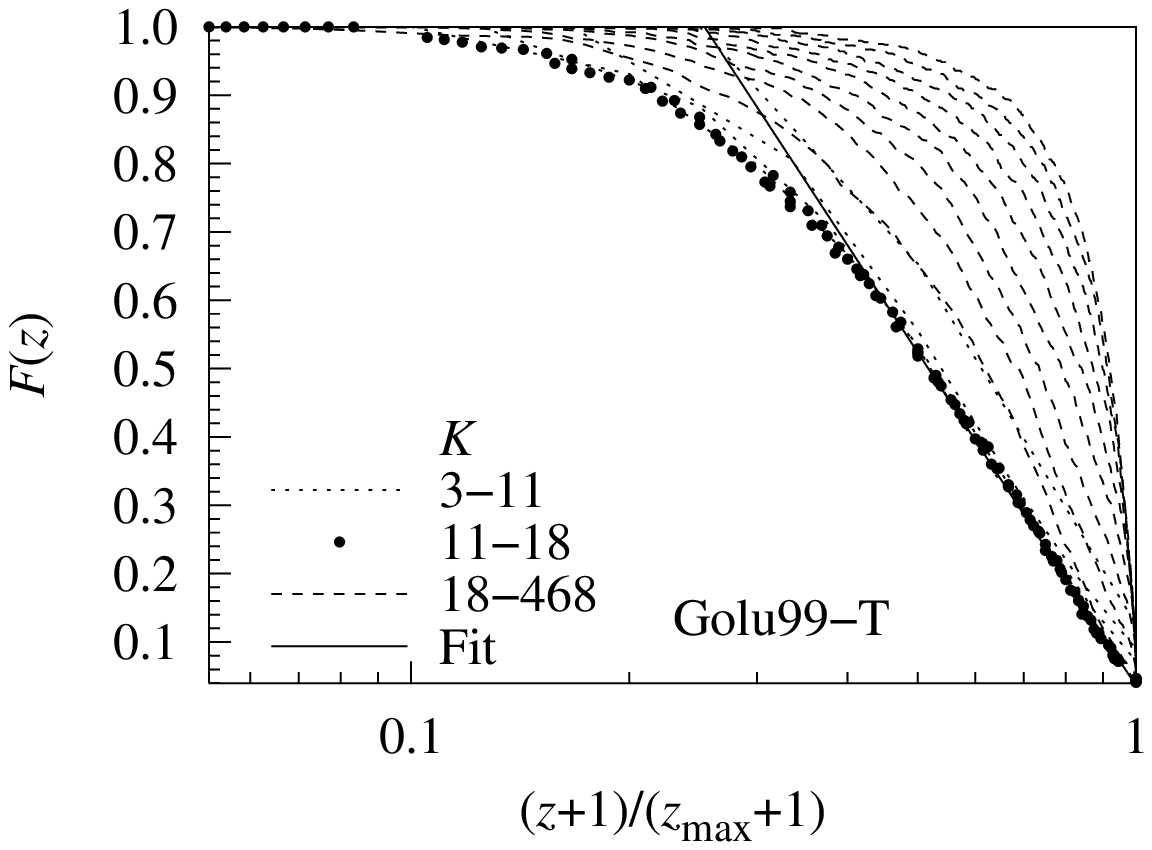}%
	    \epsfxsize55mm\epsfbox{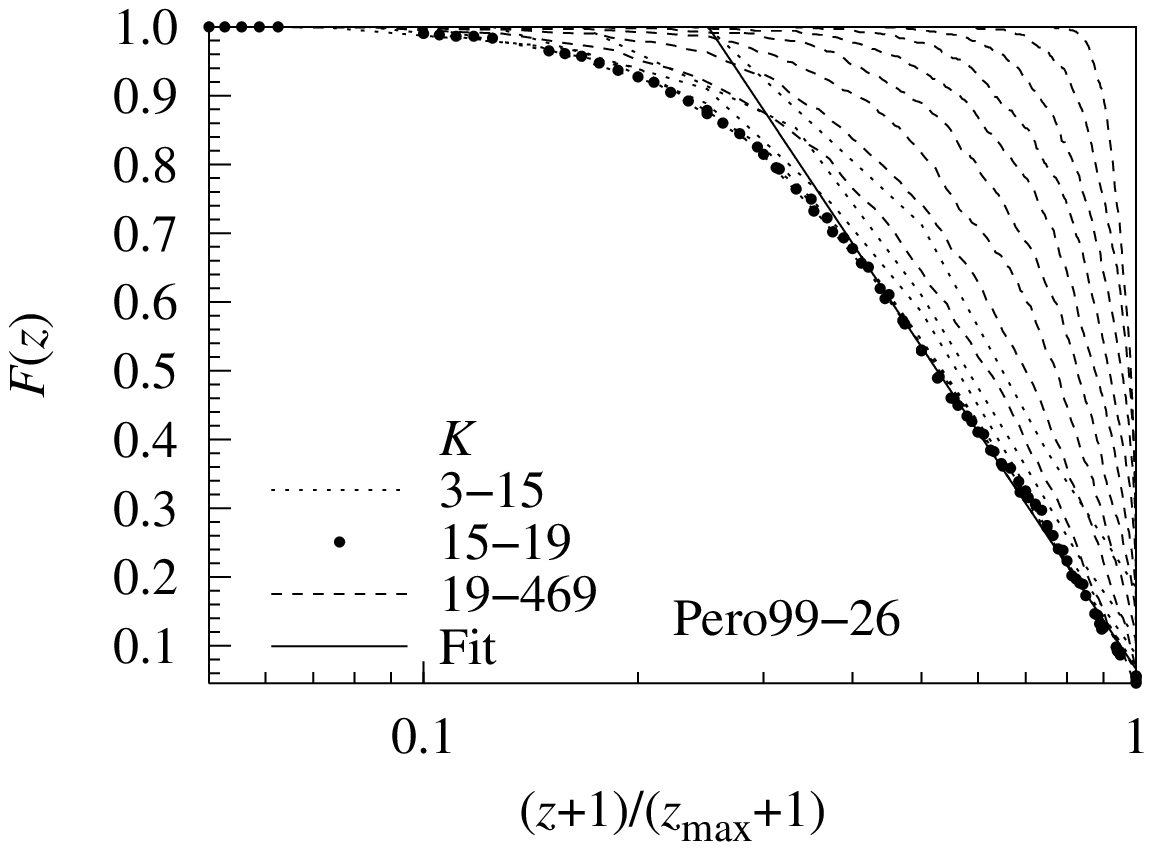}%
	    \epsfxsize55mm\epsfbox{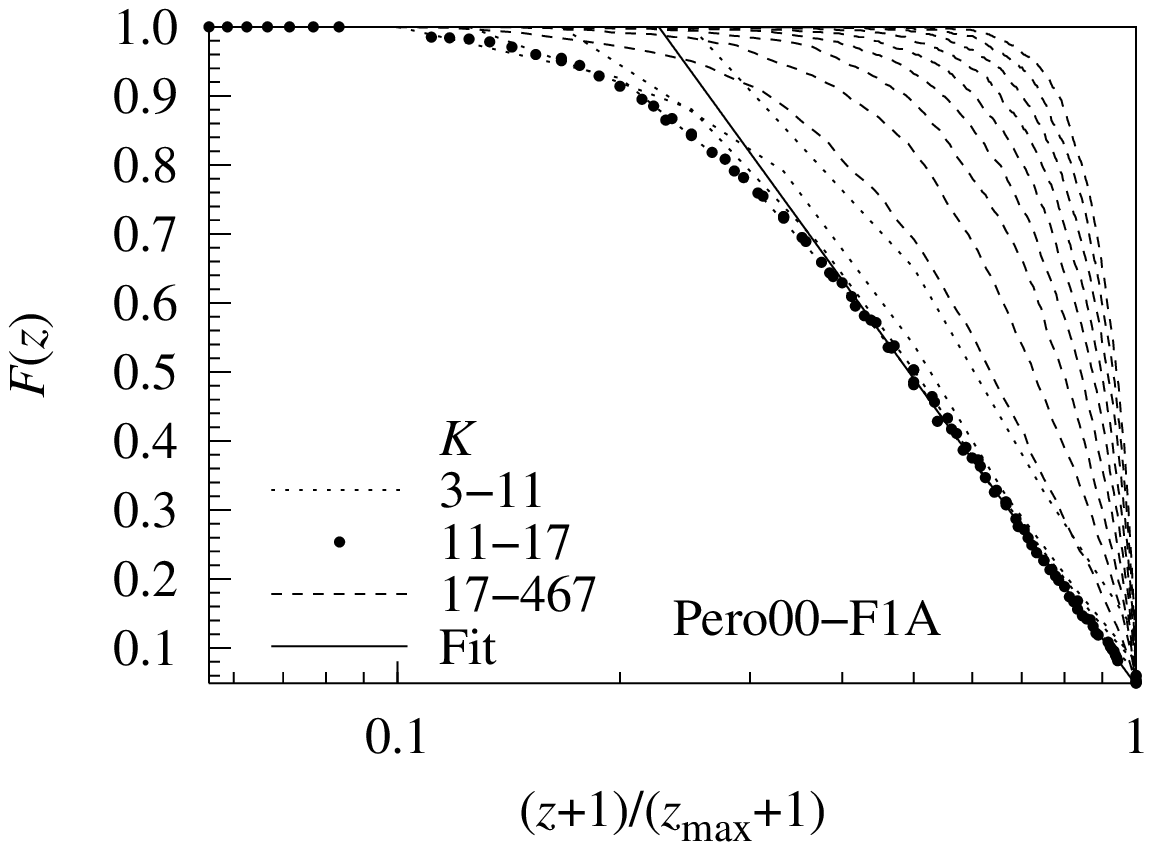}}

\caption{Variation of cumulative probability distribution function $F(z)$ with
normalized degree $(z+1)/(z_{\text{max}}+1)$ in networks constructed from
several data sets. The keys are as in \cite{DKey}. In the graphs, data for
different types of networks are plotted using different line styles. Solid
circles are used in the range of $K$ corresponding to minima of $\Lambda(K)$.
The straight solid line is least-square fit of the form given in
Eq.~(\ref{E:Fit}) in the tails of $F(z)$. The curves drawn with dashed lines
approach the solid circles as $K$ increases (here, in steps of 2). The curves
drawn with dotted lines go away from the solid circles as $K$ increases (here,
in steps of 50).}

\label{F:HistC}
\end{figure*}

Figure~\ref{F:HistC} shows the variation of the observed cumulative probability
distribution $F(z)$ with the normalized degree $(z+1)/(z_{\text{max}}+1)$ in a
wide range of values of $K$ for networks constructed from many gene expression
data sets \cite{Alon99,Nott01,Golu99,Pero99,Pero00}. It is clear from the figure
that all the curves in the range $K_{1} \le K \le K_{2}$ (solid circles) show a
very good collapse, and thus good scaling, for all the data sets. This range of
$K$ houses the minimum of the order parameter and the corresponding networks are
highly inhomogeneous with loosely connected vertices. The solid straight line
passing through the tails of these curves is a least-square fit of the form
\begin{equation}\label{E:Fit}
  F(z) = a - b \ln\left(\frac{z+1}{z_{\text{max}}+1}\right)
\end{equation}
where $a$ and $b$ are the fit parameters. For the graphs shown in the figure,
these parameters vary in the range $0.03 \le a \le 0.08$ and $0.47 \le b \le
0.71$ over all the data sets.

The extremely good fit of Eq.~(\ref{E:Fit}) in the tails of $F(z)$ seen in
Fig.~\ref{F:HistC} implies that the corresponding probability density functions
have a scale-free behavior of the form $P(z) \sim b(z+1)^{-1}$ in the tails. The
power law is seen in the range $0.35 \lesssim (z+1)/(z_{\text{max}}+1) \le
1$, i.e., in nearly 60\% to 65\% of the range of variation of vertex degree. As this
range is small for observing heavy tailed distributions, we analyze the
eigenvalue spectrum of the adjacency matrix of networks \cite{Far01Goh01} to
have another evidence of the scale-free character of the networks.

The spectral density $\rho(\lambda)$ of the eigenvalue spectrum of the adjacency
matrix of networks
\begin{equation}
  \rho(\lambda) = \frac{1}{N} \sum_{j = 1}^{N} \delta(\lambda-\lambda_{j}),
\end{equation}
where $\lambda_{j}$ is the $j$th eigenvalue of the adjacency matrix, is a good
indicator of the overall behavior of their degree distribution $P(z)$ and
topological structure. For random graphs having a giant connected component
$\rho(\lambda)$ is known to converge to a semicircle following the Wigner's law
\cite{Far01Goh01}. Deviations from Wigner's law are seen for other cases. For
the small-world networks $\rho(\lambda)$ shows a complex highly skewed structure
with several blurred peaks \cite{Far01Goh01}. For scale free networks, the
spectral density has a triangular shape with the central part lying above the
semicircle \cite{Far01Goh01}.

\begin{figure*}
\centerline{\epsfxsize51mm\epsfbox{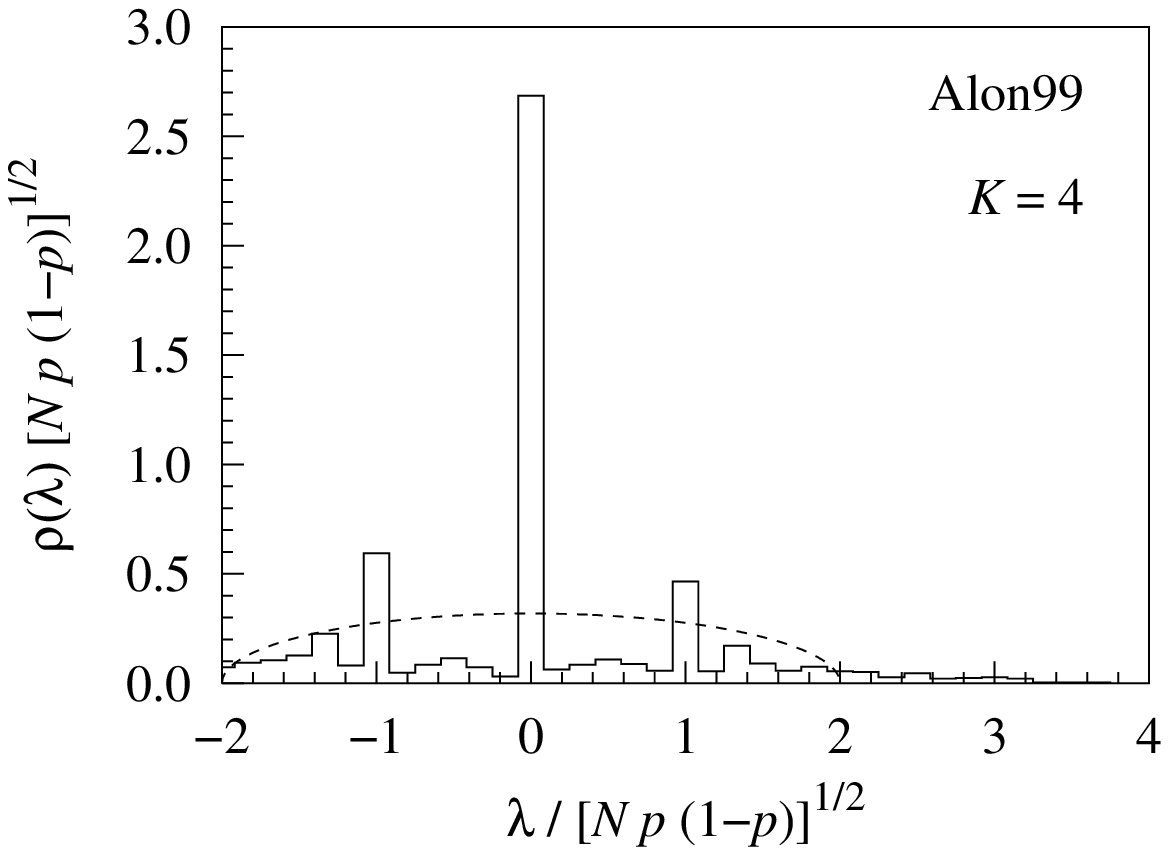}%
	    \epsfxsize51mm\epsfbox{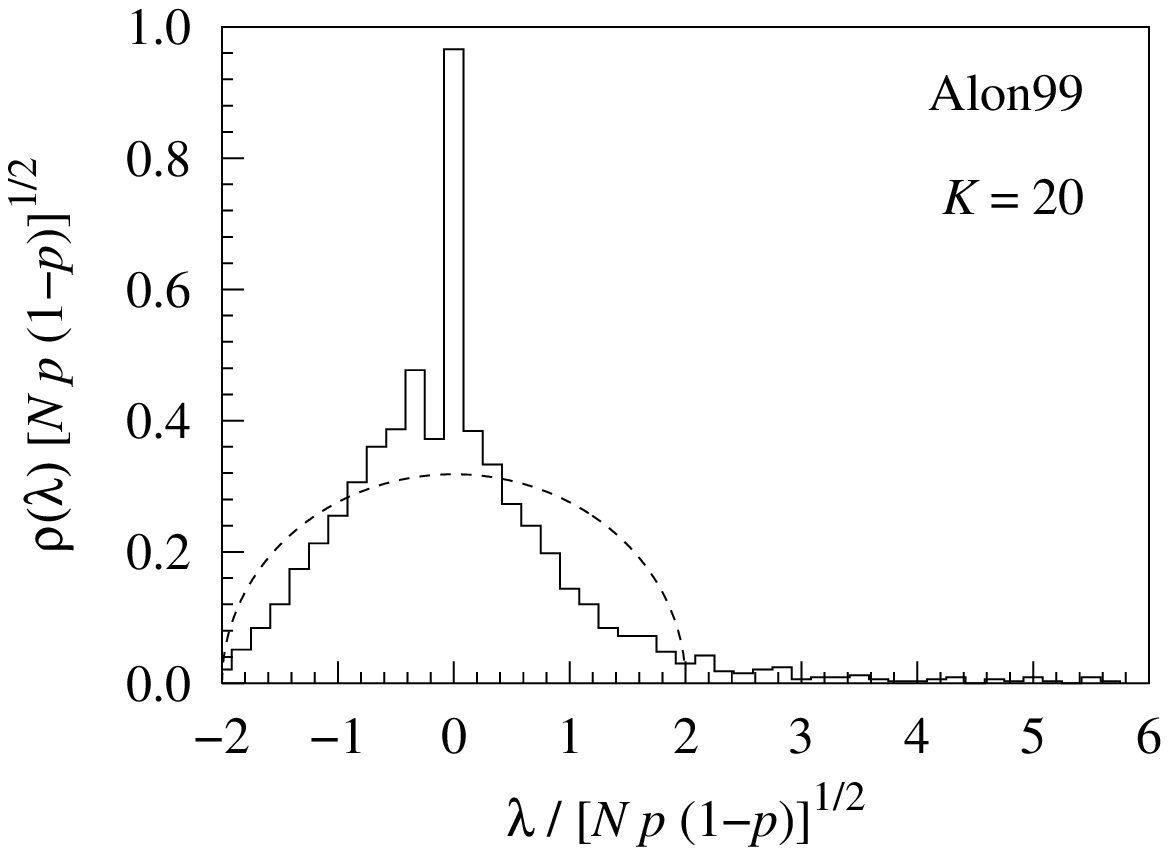}%
	    \epsfxsize51mm\epsfbox{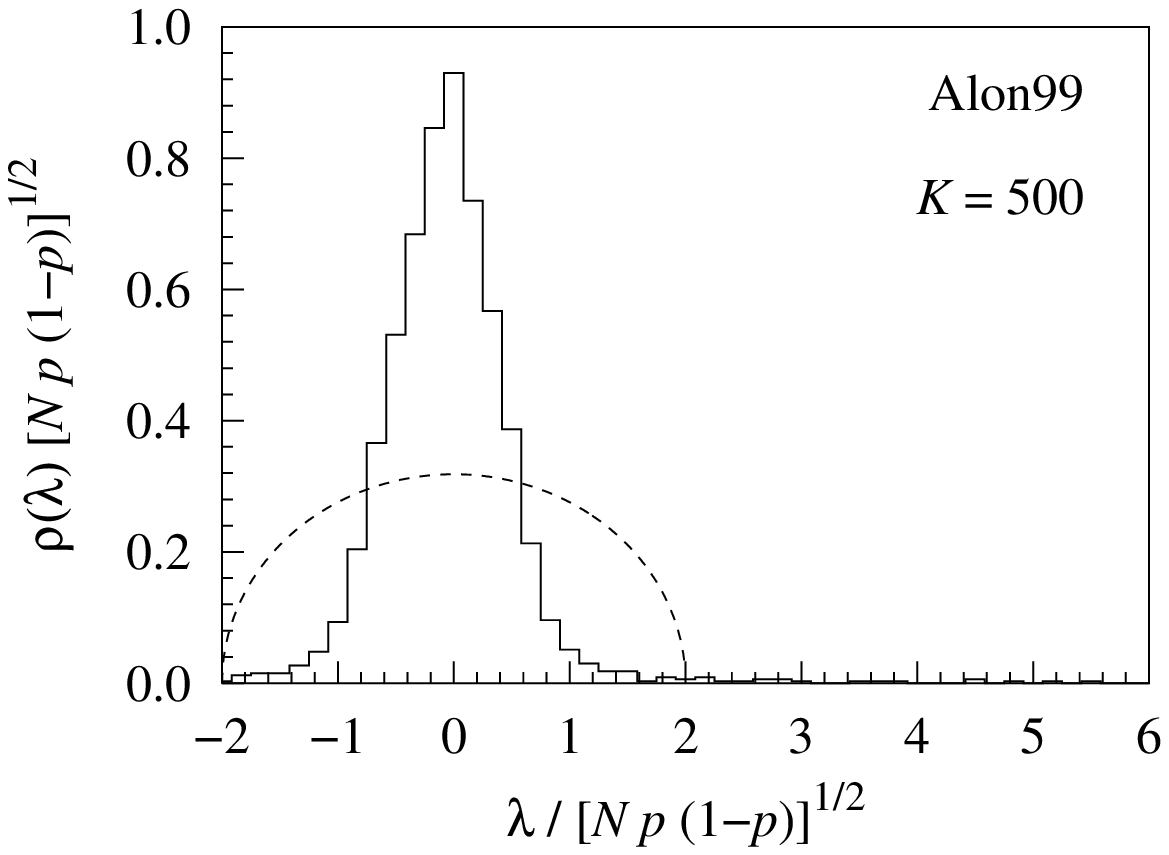}}

\caption{Spectral density for networks constructed from colon cancer data of
Alon \textit{et al.} \cite{Alon99} for different values of $K$ corresponding to
the three zones of behavior of the order parameter $\Lambda(K)$ (see
Fig.~\ref{F:LK}). $p$ is the fraction of edges, out of the maximum possible
$N(N-1)/2$, present in the network. Semicircle corresponding to spectral density
of random networks is drawn for comparison.}

\label{F:EigS}
\end{figure*}

Figure~\ref{F:EigS} shows the spectral density of networks constructed from colon
cancer data of Alon \textit{et al.} \cite {Alon99} for different values of $K$
corresponding to the three zones of behavior of $\Lambda(K)$ seen previously.
The figure shows that for small $K$ the spectral density has an irregular shape
with several blurred peaks and its bulk is confined below the semicircle. This
is a characteristic of small-world networks \cite{Far01Goh01}. As $K$ is increased,
the bulk portion starts assuming a triangular shape which is a little skewed for
small $K$. The top portion of the triangle starts protruding above the
semicircle as $K$ crosses $K_{1}$. This is clear from the intermediate and high
values of $K$ in the figure. The triangular shape persists till $K$ becomes
almost equal to $N-1$. At $K = N-1$, the spectral density develops a delta
function peak corresponding to $N-1$ repeated eigenvalues at $\lambda = -1$ and
the largest eigenvalue is $\lambda_{1} = N-1$. This behavior of the spectral
density was seen in all the other data sets also
\cite{Nott01,Golu99,Pero99,Pero00}. It confirms that these gene networks
have small-world character and become scale-free for $K \ge K_{1}$.

Presence of scale-free behavior indicates high degree of self-organization in
the system and is known to be a characteristic of natural systems \cite{Bak87}.
It has been observed in several natural and artificial networks, e.g., power
grid \cite{Wat98}, the Internet \cite{Fal99Coh00}, the World Wide Web
\cite{Alb99Hub99}, actor network \cite{Bar99}, web of human sexual contacts
\cite{Lil01}, citation and collaboration networks \cite{Red98New01}, conceptual
network of language \cite{Mot02}, metabolic network \cite{Jeo00}, food web
\cite{Sol00}, and protein networks \cite{Jeo01Mas02}.
The exponents of the power laws observed earlier, however, were always more than
unity. The gene networks studied here are first examples of biological networks
showing scale-free behavior with exponent of unity.

The scale-free character of coexpressed gene networks means that these networks
are extremely inhomogeneous and contain few genes that are very highly connected
and a large number of genes with low connectivity. This implies that these
networks contain large groups of coexpressed genes. As a result, the present
study conclusively shows, using direct experimental data, that various types of
cancers are a consequence of malfunction of only a few genes that either
regulate the expression of a large number of other genes or form the hubs of
various crowded gene regulatory pathways functioning in the organism. It is
known that disturbing such genes could be fatal for the organism \cite{Has01},
which also turns out to be the mechanism of origin of cancers. Identification
of such genes and understanding their functionality under various conditions
through which an organism can pass in its lifetime is directly relevant in the
design of highly targeted drugs, among other possibilities.

The simultaneous presence of small-world and scale-free characters in these
networks seems to be a perfect fit in the evolutionary scheme of biological
systems. The high robustness displayed by biological systems is a consequence of
the scale-free character of the associated networks. On the other hand, the fast
reaction and rapid adaptability shown by biological systems can come only if the
associated networks have a small-world character. This fits the structure of
biological signaling system well because the chemical signaling employed at most
places in biological systems, by its very nature, is very slow compared to,
e.g., electrical signaling in neurons. Thus, for achieving fast message
transmission, the associated networks must evolve to have small-world character.

I thank Eytan Domany for introducing me to gene expression data analysis and
helpful discussions, Itai Kela for help in filtering breast cancer data, and
Deepak Dhar for critical review of the manuscript.


\begin{thebibliography}{}

\bibitem{Land99Gerh99}
  E. S. Lander,
  \textrm{Nature Genet.} \textbf{21}, (1999);
  D. Gerhold, T. Rushmore, and C. T. Caskey,
  \textrm{Trends Biochem. Sci.} \textbf{24}, 168 (1999).

\bibitem{Alon99}
  U. Alon \textit{et al.},
  \textrm{Proc. Natl. Acad. Sci. U.S.A.} \textbf{96}, 6745 (1999).

\bibitem{Nott01}
  D. A. Notterman, U. Alon, A. J. Sierk, and A. J. Levine,
  \textrm{Cancer Res.} \textbf{61}, 3124 (2001).

\bibitem{Golu99}
  T. R. Golub \textit{et al.},
  \textrm{Science} \textbf{286}, 531 (1999).

\bibitem{Pero99}
  C. M. Perou \textit{et al.},
  \textrm{Proc. Natl. Acad. Sci. U.S.A.} \textbf{96}, 9212 (1999).

\bibitem{Pero00}
  C. M. Perou \textit{et al.},
  \textrm{Nature (London)} \textbf{406}, 747 (2000).

\bibitem{Braz00}
  A. Brazma and J. Vilo,
  \textrm{FEBS Lett.} \textbf{480}, 17 (2000).

\bibitem{Eis98}
  M. B. Eisen, P. T. Spellman, P. O. Brown, and D. Botstein,
  \textrm{Proc. Natl. Acad. Sci. U.S.A.} \textbf{95}, 14863 (1998).

\bibitem{Blat96}
  M. Blatt, S. Wisemann, and E. Domany,
  \textrm{Phys. Rev. Lett.} \textbf{76}, 3251 (1998).

\bibitem{Jeo01Mas02}
  H. Jeong \textit{et al.},
  \textrm{Nature (London)} \textbf{411}, 41 (2001);
  S. Maslov and K. Sneppen,
  \textrm{Science} \textbf{296}, 910 (2002).

\bibitem{Fuk90}
  K. Fukunaga,
  \textit{Introduction to statistical pattern recognition}
  (Academic, San Diego, 1990).

\bibitem{Far01Goh01}
  I. J. Farkas \textit{et al.},
  \textrm{Phys. Rev. E} \textbf{64}, 026704 (2001);
  K.-I. Goh, B. Kahng, and D. Kim,
  \textit{ibid.} \textbf{64}, 051903 (2001).

\bibitem{Bak87}
  P. Bak, C. Tang, and K. Wiesenfeld,
  \textrm{Phys. Rev. Lett.} \textbf{59}, 381 (1987).

\bibitem{Wat98}
  D. J. Watts and S. H. Strogatz,
  \textrm{Nature (London)} \textbf{393}, 440 (1998).

\bibitem{Fal99Coh00}
  M. Faloutsos, P. Faloutsos, and C. Faloutsos,
  \textrm{Comput. Commun. Rev.} \textbf{29}, 251 (1999);
  R. Cohen, K. Erez, D. ben-Avraham, and S. Havlin,
  \textrm{Phys. Rev. Lett.} \textbf{85}, 4626 (2000).

\bibitem{Alb99Hub99}
  R. Albert, H. Jeong, and A.-L. Barab\'asi,
  \textrm{Nature (London)} \textbf{401}, 130 (1999);
  B. A. Huberman and L. A. Adamic,
  \textit{ibid.} \textbf{401}, 131 (1999).

\bibitem{Bar99}
  A.-L. Barab\'asi, and R. Albert,
  \textrm{Science} \textbf{286}, 509 (1999).

\bibitem{Lil01}
  F. Liljeros \textit{et al.},
  \textrm{Nature (London)} \textbf{411}, 907 (2001).

\bibitem{Red98New01}
  S. Redner,
  \textrm{Eur. Phys. J. B} \textbf{4}, 131 (1998);
  M. E. J. Newman,
  \textrm{Proc. Natl. Acad. Sci. U.S.A.} \textbf{98}, 404 (2001).

\bibitem{Mot02}
  A. E. Motter \textit{et al.},
  \textrm{Phys. Rev. E} \textbf{65}, 065102R (2002).

\bibitem{Jeo00}
  H. Jeong \textit{et al.},
  \textrm{Nature (London)} \textbf{407}, 651 (2000).

\bibitem{Sol00}
  R. V. Sol\'e and J. Montoya,
  \textrm{Proc. R. Soc. Lond. B} \textbf{268}, 2039 (2001).

\bibitem{Has01}
  J. Hasty and J. J. Collins,
  \textrm{Nature (London)} \textbf{411}, 30 (2001).

\bibitem{DKey}
  Keys in the graphs are as follows.
  \textit{Alon99:} 2000 genes in 62 samples from colon cancer data
  \cite{Alon99}.
  \textit{Nott01-CF:} 1211 genes in 36 samples from colon adenocarcinoma data
  \cite{Nott01}. We used a nominal threshold of 20 and selected genes
  having standard deviation at least twice the mean standard deviation in
  log$_{2}$-transformed expression data.
  \textit{Golu99-I (Golu99-T):} 1049 (1030) genes in 34 (38) samples of the
  independent (training) set from acute leukemia data \cite{Golu99}. Genes
  were selected as for Nott01-CF using nominal threshold of 215 (222).
  \textit{Pero99-26:} 1030 genes in 26 samples in breast cancer data
  \cite{Pero99}. We selected genes that had good data in at least 20 samples and
  showed at least 2.5 fold variation above the median in at least two samples.
  \textit{Pero00-F1A:} 1753 genes in 84 samples from breast cancer in
  supplementary data \cite{Pero00}.

\end{thebibliography}
\end{document}